\RequirePackage[columnwise]{lineno}
\documentclass[aps,prl,twocolumn,superscriptaddress,footinbib]{revtex4} 

 \usepackage{graphicx}  
 \usepackage{dcolumn}   
 \usepackage{bm}        
 \usepackage{braket}
 \usepackage{exscale}                  
 \usepackage[intlimits]{amsmath}       
 \usepackage{amsfonts}
 \usepackage{amssymb,amscd}
 \usepackage{color}
 \usepackage{hyperref}
 \usepackage{subfigure}
 \usepackage[normalem]{ulem}
 \usepackage{slashed}
 \usepackage{leftidx}
 \usepackage{comment}
 \usepackage{booktabs}
 
 \usepackage{xcolor}
\usepackage[normalem]{ulem}

\begin{document}

\title{Realization of a Superleak in a Quantum Gas}

\author{Wan-Ling Yang}
\thanks{These authors contributed equally to this work.}
\affiliation{State Key Laboratory of Quantum Functional Materials, Department of Physics and Guangdong Basic Research Center of Excellence for Quantum Science, Southern University of Science and Technology, Shenzhen 518055, China}
\author{Shao-Bo Liu}
\thanks{These authors contributed equally to this work.}
\affiliation{State Key Laboratory of Quantum Functional Materials, Department of Physics and Guangdong Basic Research Center of Excellence for Quantum Science, Southern University of Science and Technology, Shenzhen 518055, China}
\author{Xiang Luo}
\thanks{These authors contributed equally to this work.}
\affiliation{State Key Laboratory of Quantum Functional Materials, Department of Physics and Guangdong Basic Research Center of Excellence for Quantum Science, Southern University of Science and Technology, Shenzhen 518055, China}
\affiliation{Quantum Science Center of Guangdong-Hong Kong-Macao Greater Bay Area, Shenzhen 518045, China}
\author{Chao Yang}
\affiliation{State Key Laboratory of Quantum Functional Materials, Department of Physics and Guangdong Basic Research Center of Excellence for Quantum Science, Southern University of Science and Technology, Shenzhen 518055, China}
\author{Hai-Lin Li}
\affiliation{State Key Laboratory of Quantum Functional Materials, Department of Physics and Guangdong Basic Research Center of Excellence for Quantum Science, Southern University of Science and Technology, Shenzhen 518055, China}
\author{Yu-Zhou Yang}
\affiliation{State Key Laboratory of Quantum Functional Materials, Department of Physics and Guangdong Basic Research Center of Excellence for Quantum Science, Southern University of Science and Technology, Shenzhen 518055, China}
\author{Yong-Kui Wang}
\affiliation{State Key Laboratory of Quantum Functional Materials, Department of Physics and Guangdong Basic Research Center of Excellence for Quantum Science, Southern University of Science and Technology, Shenzhen 518055, China}
\author{Bing Yang}
\email[e-mail:]{yangbing@sustech.edu.cn}
\affiliation{State Key Laboratory of Quantum Functional Materials, Department of Physics and Guangdong Basic Research Center of Excellence for Quantum Science, Southern University of Science and Technology, Shenzhen 518055, China}
\affiliation{Quantum Science Center of Guangdong-Hong Kong-Macao Greater Bay Area, Shenzhen 518045, China}



\begin{abstract}
Ultracold quantum gases offer a controllable and microscopically accessible platform for understanding and engineering superfluidity. Yet, realizing a superleak that selectively transports the superfluid while suppressing normal flow has remained elusive. Here, we harness an optical superlattice to realize such a superleak through dynamical selectivity. The superfluid undergoes coherent Josephson transport, whereas the normal component exhibits slow, thermodynamically limited response, resulting in a flow-rate contrast of 500(49). Exploiting this contrast, we demonstrate repeated filtering of the normal component and dynamical purification of the superfluid, while sustaining coherent driven oscillations for up to 100 cycles. Our results establish a controllable atomic superleak and open possibilities for manipulating transport and thermodynamics in quantum fluids.
\end{abstract}
\date{\today}
\maketitle

{\it Introduction.}
Superfluidity, a unique quantum state, fundamentally reshapes our understanding of matter~\cite{Kapitza1938,Allen1938,Landau1941} and offers a wide range of applications from condensed matter physics to quantum technologies~\cite{Leggett2006,Nielsen2012}.
Initially discovered in liquid helium~\cite{Kapitza1938,Allen1938}, 
superfluidity is characterized by the strikingly different transport responses of the superfluid and normal components.
In helium, this distinction is embodied by the \emph{superleak}, a channel that selectively transmits the superfluid component while strongly suppressing normal flow.

Ultracold atomic gases provide a highly tunable platform for exploring superfluidity and novel forms of quantum matter~\cite{Leggett2006,Pethick2008,Bloch2008,Pitaevskii2016}. 
Many hallmark phenomena of superfluid helium have found their counterparts in ultracold gases, including frictionless transport~\cite{Raman1999,Engels2007,Christodoulou2021,Desbuquois2012,Li2022,Yan2024}, quantized vortices~\cite{Matthews1999,Madison2000,AboShaeer2001}, and persistent flow~\cite{Ryu2007,Eckel2014,Polo2025}. 
In particular, micrometer-scale barriers and optical lattices have been used to realize Josephson junctions, revealing coherent tunneling, nonlinear self-trapping, and other forms of coherent superfluid dynamics~\cite{Smerzi1997,Cataliotti2001,Albiez2005,Valtolina2015,Levy2007,Kwon2020,Luick2020,Biagioni2024,Bernhart2024,Pace2024}. 
The realization of a superleak has also been proposed for ultracold gases, with disorder suggested as a means of suppressing normal transport~\cite{Papoular2014}.
However, an experimental realization of controlled superfluid-selective transport has remained elusive.

Here, we realize a dynamical superleak by exploiting the intrinsically different dynamical responses of the superfluid and normal components.
Using an optical superlattice, the superfluid undergoes coherent transport, while the response of the low-dimensional normal component is strongly suppressed by slow thermalization.
We observe a pronounced separation of their characteristic timescales, with the superfluid flow rate exceeding that of the normal component by a factor of 500(49).
A microscopic theory quantitatively captures the observed dynamics and reveals distinct scaling behaviors governed by coherent tunneling and thermalization, respectively.
Combined with site-selective removal, this dynamical selectivity enables repeated filtering of the normal component, thereby purifying the superfluid and sustaining coherent dynamics over long times.

\begin{figure*}[!htb]
\includegraphics[width=16.5cm]{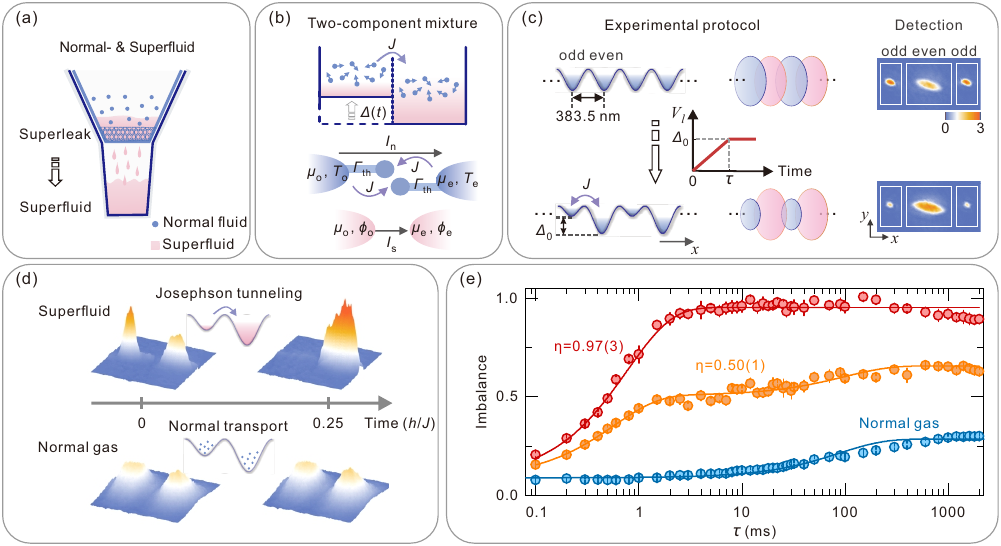}
\caption{Conceptual description of dynamical distinction between superfluid and normal transport.
(a) Schematic of a superleak in superfluid helium, where microscopic channels allow superfluid flow while suppressing the normal component.
(b) Transport in the quantum gas system. An energy bias $\Delta(t)$ drives particles through a weak link, with coherent superfluid tunneling and suppressed normal transport.
(c) Experimental protocol. The energy offset is linearly ramped over a time $\tau$, and site-resolved populations are measured by band-mapping and time-of-flight (TOF) imaging.
(d) Transport in a double-well unit of the optical superlattice. After ramping $\Delta$, the superfluid rapidly transfers into the deeper well within $h/4J$, while the normal component shows strongly suppressed transport.
(e) Transport dynamics. Atom-number imbalances are shown on a logarithmic time scale, with the condensate fraction $\eta$ indicated. The nearly pure condensate and thermal samples are fitted by single-exponential decays, while the mixed sample with a condensate fraction of 50(1)\% is described by a double-exponential fit.}
\label{Fig1}
\end{figure*}

\begin{figure*}[!htb]
\includegraphics[width=17.1cm]{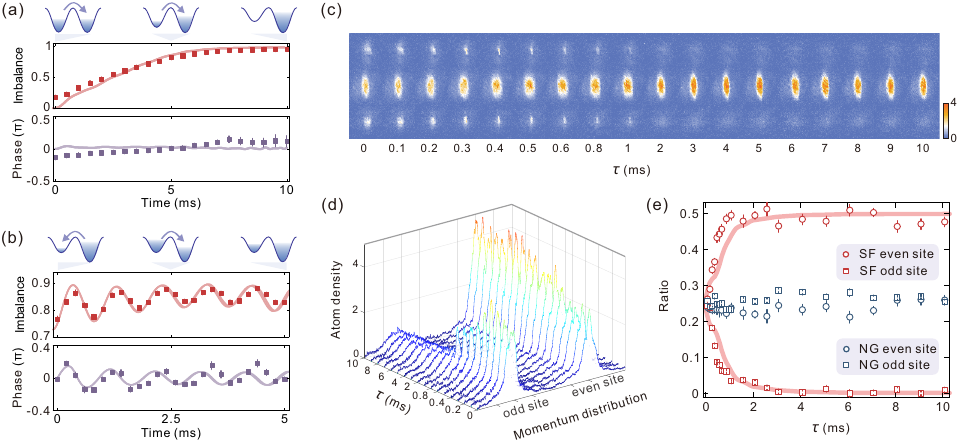}
\caption{
Superfluid phase coherence and component-selective transport.
(a) Superfluid transport via adiabatic ramping. The phase difference between odd and even sites is measured during transport. Data points show the experimental results with standard deviations, and solid lines are theoretical predictions.
(b) Superfluid Josephson oscillation. Following a sudden quench, both the site population imbalance and relative phase exhibit coherent oscillations, in agreement with theoretical predictions.
(c) Time-resolved transport dynamics. The atomic distribution of a condensate-thermal mixture is monitored during an energy-offset ramp with duration $\tau$.
(d) Site-resolved density evolution. The density profiles extracted from band-mapping images are fitted with a bimodal function to obtain the condensate and thermal components.
(e) Spatial separation of the two components. The time evolution of the coherent and thermal populations on even and odd sites is shown. Red (blue) markers represent the superfluid (thermal) response, with circles (squares) indicating even (odd) sites.
}
\label{Fig2}
\end{figure*}


{\it Theoretical descriptions.}
For a superleak in liquid helium, a porous material with sufficiently narrow channels suppresses viscous normal flow while allowing the inviscid superfluid to pass (see Fig.~\ref{Fig1}(a)). 
Such material-based filtering is not available in ultracold gases. 
Instead, we engineer a dynamical superleak using an optical superlattice, where neighboring quasi-two-dimensional (2D) regions are coupled through optical weak links, as shown in Fig.~\ref{Fig1}(b), (c).
In this bichromatic superlattice~\cite{Bloch2008,Yang2020},
neighboring odd and even subsystems experience a time-dependent energy offset $\Delta(t)$, which drives particle redistribution between the two sublattices. 
The transport dynamics of each component are characterized by the flow rate
$I_{\text{s},\text{n}}=\partial_t(N_{s,n}^{\rm o}-N_{s,n}^{\rm e})/2$, where
$N_{s,n}^{\rm o},~N_{s,n}^{\rm e}$ denote the population of the superfluid (s) or normal (n) component in odd (o) and even (e) sites. 

For the superfluid component, macroscopic phase coherence gives rise to phase rigidity, with the superfluid velocity determined by the spatial phase gradient~\cite{Pethick2008,Pitaevskii2016},
\begin{equation}
\label{eq:superfluid_velocity}
\mathbf{v}_{s}=\frac{\hbar}{m}\nabla\phi,
\end{equation}
where $\hbar$ is the reduced Planck constant, $m$ is the atomic mass, and $\phi$ is the superfluid phase.
The corresponding particle-current density is given by $\mathbf{j}_s=n_s\mathbf{v}_s$, where $n_s$ denotes the superfluid number density.
In this lattice system, the phase difference $\Delta\phi$ between neighboring wells gives rise to Josephson coupling, described by the current-phase relation $I_s=I_{\mathrm{c}}\sin(\Delta\phi)$~\cite{Smerzi1997}, where $I_s$ is the Josephson current per particle and $I_{\mathrm{c}}$ is its critical value.
The chemical-potential difference $\Delta \mu = \mu_{\rm o}- \mu_{\rm e}$ between adjacent wells drives the evolution of $\Delta\phi$ and can generate Josephson oscillations or self-trapping~\cite{Albiez2005}.
Here, the energy offset $\Delta(t)$ between the odd and even wells is adiabatically modulated to drive controlled superfluid transport while suppressing oscillatory Josephson dynamics.

In contrast, the normal component lacks phase coherence and its transport is governed by incoherent quantum tunneling. 
Initially, the odd and even sublattices are thermodynamically balanced, with identical bosonic occupation distributions $f_{\rm o}(E)=f_{\rm e}(E)$, resulting in no net normal current. 
When an energy offset $\Delta$ is applied, atoms tend to redistribute along the lattice direction ($x$) toward lower-energy sites.
Establishing the corresponding thermal equilibrium, however, requires the energy released along the lattice direction to be redistributed into the in-plane thermal degrees of freedom.
This process relies on cross-dimensional thermalization between the axial lattice motion and the transverse $y$-$z$ degrees of freedom.
We denote the corresponding thermalization rate by $\Gamma_{\rm th}$.
In-plane collisions dephase inter-site tunneling and redistribute the transferred energy. 
In the tight-binding regime, this cross-dimensional thermalization is strongly suppressed~\cite{Petrov2001}, resulting in a delayed buildup of the occupation imbalance $f_{\rm o}(E)-f_{\rm e}(E)$ between neighboring sublattices.

Once the occupation imbalance is established, the normal transport can be described by a Landauer-type relation,
\begin{equation}
\label{eq:landauer}
I_n=\frac{1}{h}\int dE\,\mathcal{T}(E)
[f_{\rm o}(E)-f_{\rm e}(E)],
\end{equation}
where $\mathcal{T}(E)$ is the transmission probability between neighboring sites and $f_{\rm o,e}(E)$ are the energy-dependent occupation distributions.
We denote the normal transport relaxation rate by $\Gamma$, which in our thermalization-limited regime scales as $\Gamma\propto\Gamma_{\rm th}$.
The slow thermalization process therefore sets the slow timescale of normal transport, leading to a delayed and dissipative response compared with the phase-coherent superfluid flow.

Together, the optical superlattice exploits the intrinsic difference between coherent superfluid flow and thermalization-limited normal transport, producing a strong component-selective transport contrast and realizing the functionality of a dynamical superleak.

{\it Experiment.} 
The experiment starts with ultracold gases of $\simeq 2 \times 10^5$ $^{87}$Rb atoms confined in a 3D optical dipole trap.
The typical trapping frequencies are $\left(\omega_x, \omega_y, \omega_z \right) = 2\pi \times \left[189(3), 65(1), 63(1)\right]~\text{Hz}$, giving the condensate a Thomas-Fermi radius of approximately $3.7~\mu$m along $x$.
The condensate fraction can be finely tuned by controlling the final depth of the dipole potential.
A bichromatic lattice along the $x$-axis divides the system into $\simeq 30$ layers of quasi-2D quantum gases, as shown in Fig.~\ref{Fig1}(c).
The wavelength of the ``short'' lattice laser is $\lambda_s = 767$ nm, while the ``long'' lattice has a wavelength of $\lambda_l=2\lambda_s$, forming an optical superlattice that categorizes the trapped atoms into odd and even subsystems.
The lattice depths are given in units of the recoil energy $E_r = h^2/(2 m \lambda_s^2)$, where $h = 2\pi \hbar$, and $m$ is the atomic mass.
The superlattice potential can be expressed as $V(x) = V_s \cos^2(kx) - V_l \cos^2(kx/2 + \varphi)$, with $k = 2\pi/\lambda_s$ and $ V_l = \Delta $.
The phase of the optical lattices, $\varphi$, is set to $\pi/4$ to align the intensity minima of the lattices and ensure identical tunneling strength between lattice sites.

We characterize the transport of the mixture by resolving the superfluid and thermal components from their momentum distributions. 
Band mapping separates the odd and even sites into different bands, and their momentum distributions are detected after a 25-ms time of flight (see Fig.~\ref{Fig1}(c)).
The population imbalance between odd and even sites can be written as, $\mathcal{I}=(N_{\rm e}-N_{\rm o})/(N_{\rm e}+N_{\rm o})$. 
The condensate fraction is extracted from a bimodal fit to the momentum distribution along the $y$ direction. 
Because the lattice only modifies the dispersion along the lattice direction ($x$) and leaves the transverse ($y$-$z$) momentum distribution unchanged, this condensate fraction provides an experimental measure of the coherently transporting component in our regime~\cite{Chauveau2023,Tao2023} (see Supplementary Material).

We induce local transport by linearly ramping the long-lattice energy offset $\Delta(t)$, with $V_s = 9.0(1)E_{\rm r}$, which drives the quasi-2D gases into neighboring deeper wells through the weak links. By preparing samples with different condensate fractions, we reveal distinct transport timescales (Fig.~\ref{Fig1}(d),(e)). 
For a nearly pure condensate with a condensate fraction of 97(3)\%, the superfluid component transfers to the neighboring wells within milliseconds, whereas a thermal sample exhibits transport over hundreds of milliseconds. 
For an intermediate-temperature sample with a condensate fraction of 50(1)\%, a clear two-stage dynamics emerges, consisting of a fast superfluid response followed by a slow thermal transport.
The amplitude of the fast coherent response closely follows the condensate fraction over the explored temperature range.
The extracted superfluid transport rate exceeds the normal component by a factor of 500(49).

\begin{figure*}[tbh]
\includegraphics[width=17cm]{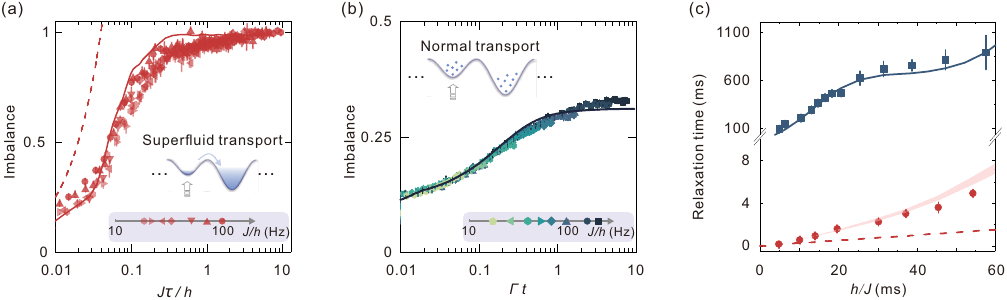}
\caption{Scaling dynamics of superfluid and normal transport.
(a) Scaled dynamics of superfluid transport. The evolution of the superfluid population imbalance is plotted as a function of the rescaled time. The tunneling rate $J/h$ is varied from $18.4(3)$~Hz to $98.3(9)$~Hz. The dashed line indicates the critical Josephson timescale.
(b) Scaled dynamics of normal transport. The evolution of the normal-component imbalance is plotted using the thermalization-related scaling parameter $\Gamma$. The solid line shows the theoretical simulation.
(c) Comparison of the characteristic timescales. The extracted relaxation times are compared with the Josephson and thermalization timescales for the superfluid and normal components, respectively. The red shaded region and blue line show the corresponding theoretical results.
}
\label{Fig3}
\end{figure*}


To establish the phase-coherent nature of the observed superfluid transport, we simultaneously measure the population imbalance and the relative phase between neighboring sites.
With $J/h=96.0~\mathrm{Hz}$, the energy offset $\Delta$ is linearly ramped over $10~\mathrm{ms}$, driving atoms toward the deeper sites.
The relative phase, extracted from the interference pattern after TOF expansion, develops a finite deviation from zero during transport (see Fig.~\ref{Fig2}(a)).
The small phase difference is consistent with the near-adiabatic ramp and accompanies the observed particle current.
Numerical simulations reproduce both the imbalance and phase evolution, supporting the phase-current relation of the transport.

We further demonstrate superfluidity by probing Josephson dynamics following a sudden quench.
The condensate is adiabatically prepared in the same lattice configuration with $V_l=0.50(1)~E_r$ and an initial population imbalance of $0.78$, after which the energy offset is quenched to $\Delta=0.63(1)~E_r$ within $0.05~\mathrm{ms}$.
Both the population imbalance and relative phase exhibit clear coupled oscillations with weak damping (Fig.~\ref{Fig2}(b)).
The two oscillations are shifted by approximately $\pi/2$, characteristic of the conjugate population and phase dynamics of a Josephson junction~\cite{Albiez2005,Smerzi1997}.
The measured oscillation frequency of $992(5)~\mathrm{Hz}$ agrees with numerical simulations, providing direct phase-sensitive evidence for coherent Josephson dynamics and the superfluid nature of the fast response.


\begin{figure*}
\includegraphics[width=16.6cm]{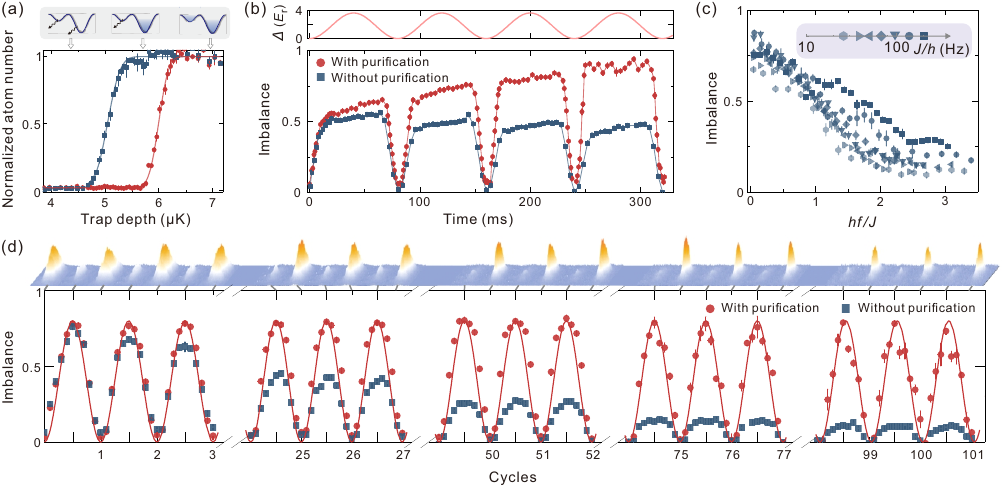}
\caption{Superleak and sustained coherent dynamics.
(a) Purification method. Site-selective removal of atoms is achieved by lowering the transverse trap depth. Red and blue points denote the normalized atom numbers on the odd and even sites, respectively. At a transverse depth of $\simeq 5.5~\mu\mathrm{K}$, atoms on the odd sites are effectively removed.
(b) Cyclic purification. The long-lattice offset $\Delta$ is sinusoidally modulated with a period of $80~\mathrm{ms}$, with atoms on the odd sites selectively removed during each cycle. Blue points show the dynamics without purification.
(c) Transport-induced excitations. 
The superfluid oscillation amplitude after $20$ driving cycles decreases with the normalized driving frequency $hf/J$. Data at different tunneling strengths nearly collapse onto a common curve.
(d) Sustained coherent dynamics. Cyclic purification maintains the driven oscillations over $100$ cycles. The upper panel shows the normalized odd-site densities, while blue points show the dynamics without purification. The solid line is a guide to the eye.
}
\label{Fig4}
\end{figure*}

The distinct transport dynamics of the superfluid and thermal components enable their spatial separation. 
We prepare a two-component mixture with a condensate fraction of 50(1)\% and drive transport by linearly ramping the long-lattice energy offset $\Delta(t)$ with a total ramp duration $\tau$. 
The final offset is chosen as $\Delta_0=1.25(1)E_r$, such that the particles are transferred toward the deeper lattice sites through the weak links. 
As shown in Fig.~\ref{Fig2}(c), the condensate component rapidly transfers from odd to even sites within approximately $1$ ms, whereas the thermal component remains nearly unchanged over a $10$-ms ramp.
The density evolution in Fig.~\ref{Fig2}(d) directly reveals the spatial separation of the coherent and thermal components. 
Quantitative analysis in Fig.~\ref{Fig2}(e) shows that the characteristic superfluid transport occurs on a timescale of approximately $0.1h/J$.

We further investigate the scaling behavior of the superfluid dynamics by varying the tunneling strength $J$. 
For a nearly pure condensate in the tight-binding regime, all transport curves collapse when the evolution time is rescaled by $J\tau/h$ (Fig.~\ref{Fig3}(a)), demonstrating that the coherent transport timescale is determined by the tunneling energy. 
The characteristic timescale of the superfluid transport is set by the Josephson critical current, with $I_c=2J/\hbar$ defining the upper bound for coherent flow through the weak link~\cite{Smerzi2002}. 
The dashed line in Fig.~\ref{Fig3}(a) indicates this critical-current limit, below which the observed current remains due to the near-adiabatic ramping process.

For comparison, we measure the dynamics of the thermal component in the same lattice configurations (see Fig.~\ref{Fig3}(b)). 
The relaxation times are extracted from single-exponential fits of the population imbalance and are approximately two orders of magnitude longer than the superfluid transport timescale. 
At the largest tunneling strength explored, $J/h=206(2)$~Hz, the superfluid transport rate exceeds the thermal transport rate by a factor of $5791(601)$.
In quasi-2D systems, energy exchange between the axial lattice motion and the in-plane thermal degrees of freedom is strongly suppressed and follows $\Gamma_{\rm th}\propto \exp(-\hbar\omega_l/k_BT)$~\cite{Petrov2001}, where $\omega_l$ is the axial trapping frequency. 
Since the tunneling-associated process is much faster, cross-dimensional thermalization sets the slow timescale of normal transport.
When the evolution time is rescaled by the effective transport rate $\Gamma$, the normal transport dynamics collapse onto a single curve, revealing the underlying scaling law of thermal transport.
This behavior is distinct from second sound, where the superfluid and normal components form a coupled hydrodynamic mode.

{\it Dynamical superleak.}
The strong transport contrast between the two components enables a dynamical filtering process analogous to a conventional superleak. 
During transport, the superfluid preferentially flows from the odd to the deeper even sites, while the thermal atoms remain nearly stationary on both sublattices. 
We then selectively remove atoms from the odd sites by exploiting the different transverse confinements of the two subsystems. 
Specifically, the red-detuned long lattice provides additional confinement for the even sites in the $y$-$z$ plane, allowing atoms on the odd sites to spill out when the transverse dipole trap is lowered, while leaving the even-site atoms largely unaffected (Fig.~\ref{Fig4}(a)).
Repeating this transport-and-removal cycle progressively filters out the normal component and purifies the superfluid.
Unlike conventional evaporative cooling, the selectivity here is established by the distinct many-body transport and thermodynamic responses of the two components before the removal step.

We implement the superleak cyclically by modulating the long lattice, as shown in Fig.~\ref{Fig4}(b). During each cycle, the superfluid is transported to the deeper sites, followed by selective removal of atoms from the odd sites. 
Starting from approximately equal superfluid and normal populations, the first cycle removes about half of the normal component. 
At finite temperature, the broadened evaporation curves in Fig.~\ref{Fig4}(a) reduce the site selectivity, resulting in some accompanying superfluid loss. 
After four filtering cycles, the condensate fraction increases from approximately 50\% to 94(1)\%, while 57\% of the initial superfluid remain. Here, the removal strength is chosen to obtain a nearly pure superfluid within only a few cycles; using weaker removal over more cycles could further reduce the superfluid loss. 
This repeated transport-and-filtering process realizes the functionality of a superleak in the time domain, preferentially retaining the superfluid while removing the normal component.

The transport also reveals the critical flow of the superfluid, as shown in Fig.~\ref{Fig4}(c). 
By varying the total ramp duration $\tau$, we control the flow rate through the weak links, with shorter ramps requiring faster superfluid transport. 
As the flow approaches the Josephson critical current \(I_c=2J/\hbar\), strong excitations emerge~\cite{Smerzi2002}. 
When the driving timescale is rescaled by \(h/J\), measurements at different tunneling strengths collapse onto a common curve, identifying the characteristic scale for the onset of excitations~\cite{Burger2001,Wu2001,Smerzi2002,Morsch2006}.

The filtering functionality further allows us to stabilize superfluid transport against the excitations accumulated during repeated driving. Without filtering, the generated thermal component progressively reduces the superfluid fraction and damps the Josephson dynamics. 
We therefore continuously apply the purification protocol during sinusoidal driving, removing thermal excitations while retaining the transported superfluid. 
As shown in Fig.~\ref{Fig4}(d), the resulting superflow remains coherent with no detectable decay over 100 driving cycles, in contrast to the pronounced damping observed without purification. 
Beyond demonstrating a dynamical superleak, this ability to selectively remove entropy-carrying excitations provides a route to maintaining low-entropy quantum matter during driven dynamics and may offer a general tool for stabilizing coherent many-body transport.

{\it Conclusions and discussion.}
In summary, we realize a dynamical superleak in an ultracold atomic gas by exploiting the distinct transport mechanisms of the superfluid and normal components. 
Coherent superfluid transport and thermalization-limited normal transport exhibit distinct dynamical scalings, resulting in a strong transport contrast. 
Combined with site-selective removal, this dynamical selectivity enables repeated filtering of the normal component, thereby purifying the superfluid and stabilizing coherent dynamics against excitations. 
As the filtering relies on thermodynamic response, it may extend to a broad class of quantum-fluid systems.

These results provide a new approach to studying and controlling and controlling superfluidity in low-dimensional systems~\cite{Hadzibabic2006,Desbuquois2012,Christodoulou2021,Chauveau2023,Tao2023}, while offering a route toward preparing high-purity superfluids in strongly correlated lattice systems~\cite{Greiner2002,Bloch2012}.
The ability to maintain superfluid coherence through dynamical filtering may benefit atomtronic quantum devices~\cite{Eckel2014,Amico2021,Amico2022} and precision measurements~\cite{Schumm2005,Esteve2008,Pezze2018,Polo2025,Zhang2026}.
More broadly, this approach provides a platform for engineering quantum fluids and exploring nonequilibrium many-body dynamics and dissipation~\cite{Leggett2006}.

\vspace*{0.5cm}

{\textbf{Acknowledgments:}}
We thank P. Hauke, S. Jochim, Q. Niu, A. Smerzi, S. Stringari, Y.-C. Wang, B. Wu, S.-Z. Zhang, H.-Z. Zhao, Q. Zhou for insightful discussions.
We thank Z.-B. Yang and J.-Y. Shang for their assistance in the initial studies.
This work is supported by the National Key R\&D Program of China (Grant No. 2022YFA1405800), NSFC (Grant No. 12274199), Shenzhen Science and Technology Program (Grant No. KQTD20240729102026004), Guangdong Major Project of  Basic and Applied Basic Research (Grant No. 2023B0303000011) and Guangdong Provincial Quantum Science Strategic Initiative (Grant No. GDZX2304006, Grant No. GDZX2405006). X. L. acknowledges China postdoctoral science foundation (No. 2023M741547).

\bibliographystyle{apsrev4-2}
\bibliography{SFpurify}


\clearpage
\onecolumngrid
\vspace*{0.5cm}
\begin{center}
    \textbf{SUPPLEMENTARY MATERIALS}
\end{center}
\vspace*{0.5cm}
\twocolumngrid
\setcounter{equation}{0}
\setcounter{figure}{0}
\makeatletter
\makeatother
\renewcommand{\theequation}{S\arabic{equation}}
\renewcommand{\thefigure}{S\arabic{figure}}

\section{Experimental system}

We prepare the cold atomic sample through the typical atom cooling procedure.
Using a crossed dipole trap, we can create a nearly pure condensate of $^{87}$Rb with 1.3$\times10^5$ atoms.
The temperature of the cloud is controlled by adjusting the depth of the dipole potential during the final stages of evaporation.
As the condensate fraction approaches 0, the total atom number increases to approximately 5$\times10^5$.
We determine the condensate fraction by fitting the time-of-flight cloud using a bimodal distribution.
The trap frequencies are calibrated through dipole oscillation measurements.
The atomic cloud has a lifetime of about 15 seconds, enabling state manipulations without significant loss.

We manipulate atoms and characterize superfluidity using an optical superlattice.
The optical superlattice consists of two overlapping lattices generated by retro-reflecting lasers.
One lattice has a wavelength of 767 nm, generated by frequency doubling from the same source as the 1534 nm lattice laser, ensuring their wavelengths relate as $\lambda_l = 2\lambda_s$.
The dipole potential of the superlattice is described by $V(x) = V_{s}$ cos$^{2}(kx) - V_{l}$ cos$^{2}(kx/2+\varphi)$, where $\varphi$ is the phase of the superlattice.
In the 179 mm retro-reflective beam path, we insert a glass plate to create a small refractive index difference, thereby adjusting the phase $\varphi$.
Additionally, we fine-tune $\varphi$ by slightly adjusting the frequency of the 1534 nm laser, altering the relative optical path.
Through an acousto-optic modulator, we change the 1534 nm laser frequency by 220 MHz, tuning $\varphi$ by $0.26\pi$.
We calibrate $\varphi$ by measuring the ratio of atom numbers between the odd and even subsystems.
Typically, we set the $\varphi = \pi$/4 in most experiments to create an imbalanced lattice potential.

The optical lattice divides atoms into layers of two-dimensional (2D) systems.
The coupling between lattice sites is characterized by the tunneling strength $J$, primarily controlled by varying the trap depth of the short lattice $V_s$.
The transversal trapping frequency is $\omega_{\bot}$ is 2$\pi\times$ 64(1) Hz, and along the $x$ direction, the lattice imposes $\omega_{l}$ = 2$\pi\times$ 23 kHz [$V_s = 9.0(1) E_r$].
The chemical potential of the 2D system is derived as,
$\mu_j = \Bigl[ m\omega_{\bot}^{2} g N_j / (\sqrt{2\pi} \pi \sigma_{l}) \Bigr]^{1/2}$, where $g = 4\pi\hbar^{2}a_{s}/m$, $a_s$ is the s-wave scattering length of $^{87}$Rb, and $\sigma_{l} = \sqrt{\hbar/m\omega_l}$.
When a BEC is loaded into this optical lattice, the atom number in the central maximum, $N_{c}$, is approximately 1.9$\times10^4$.
Consequently, the chemical potential in the central part of the cloud is $\mu_{c}$ $\simeq k_B\times 150$ nK.

\begin{figure}[t]
\centering
\includegraphics[width=7.5cm]{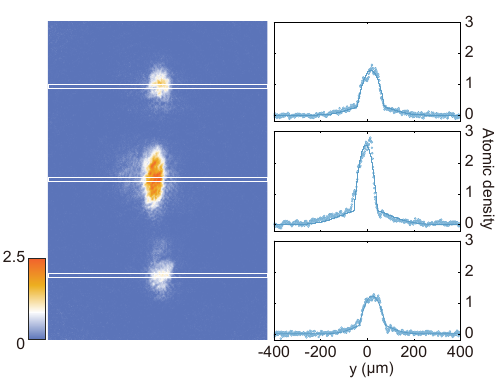}
\caption{Detecting the condensate fraction.
In the left panel, white rectangular areas are selected to extract the density distribution along the $y$-direction. The atomic density distributions are displayed in the right panel. Experimental data is represented by blue dots, while the solid lines depict the results of bimodal fitting.}
\label{FigS1}
\end{figure}

To measure the properties of the odd and even subsystems, we employ a site-resolved band-mapping method to spatially separate the subsystems and then detect their density distributions.
During the band-mapping process, atoms in the odd subsystem are transferred to the higher band of the optical lattice by increasing the potential $V_l$ to $20~E_r$.
Simultaneously, the short lattice is raised to $30~E_r$ to enlarge the band gap.
Following this, the short lattice is abruptly switched off within  $300~\mu$s, merging two neighboring sites into a long-lattice site and transferring atoms of the odd subsystem to the higher band.
Subsequently, we ramp down the long lattice over 700 $\mu$s and allow the cloud to expand freely for 25 ms before capturing absorption images.
The typical image is depicted in Fig. \ref{FigS1}, where the even (odd) subsystem is mapped onto the main (side) peak(s).
Since the band-mapping process is conducted along the $x$-direction, the transversal momentum distribution along $y$ or $z$ reflects the original 2D cloud properties.

\begin{figure}[htbp]
\includegraphics[width=7.8cm]{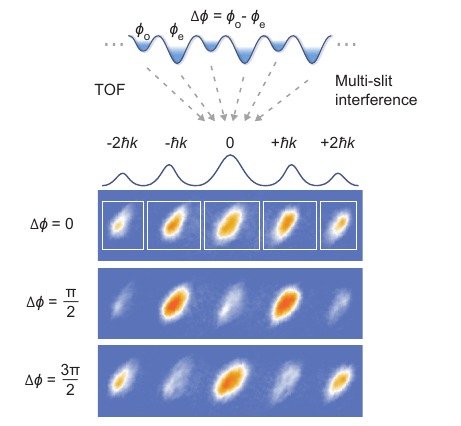}
\caption{Measurement of the relative phase.
Representative TOF interference patterns between the odd and even subsystems.
Starting from the initial state prepared at $\Delta=0.50(1)~E_{r}$, matter-wave interference produces characteristic intensities at the diffraction orders $0$, $\pm\hbar k$, and $\pm2\hbar k$ that depend on the relative phase $\Delta\phi$, allowing $\Delta\phi$ to be extracted from the interference pattern.}
\label{FigS12}
\end{figure}

The superfluid dynamics are characterized by the population imbalance and the relative phase $\Delta\phi$ between the odd and even subsystems.
We measure $\Delta\phi$ from the interference pattern after time-of-flight (TOF) expansion, during which matter waves from the two subsystems overlap.
Since the system consists of an array of coherent double-well units, the resulting pattern resembles multi-slit interference and contains diffraction peaks at integer multiples of $\hbar k$.
For equal populations in the odd and even sites and $\Delta\phi=0$, the interference is dominated by the short-lattice diffraction peaks at $\pm 2\hbar k$, whereas a finite $\Delta\phi$ transfers population to the $\pm\hbar k$ peaks.
The populations of these diffraction orders therefore provide a sensitive probe of the relative phase.

To calibrate the phase measurement, we prepare samples with different relative phases and record the corresponding diffraction patterns. Figure~\ref{FigS12} shows a representative calibration measurement at $\Delta=0.50(1)~E_{r}$.
For each TOF image, we integrate the populations of the five main diffraction peaks and use the population fraction in the $\pm\hbar k$ peaks as the interference signal.
Combining this signal with the measured population imbalance and the calibrated phase dependence allows us to extract $\Delta\phi$.
Repeated measurements at different evolution times then yield the phase dynamics between the odd and even subsystems.

\section{Coherent-fraction calibration}

\begin{figure}[tbp]
\includegraphics[width=7.8cm]{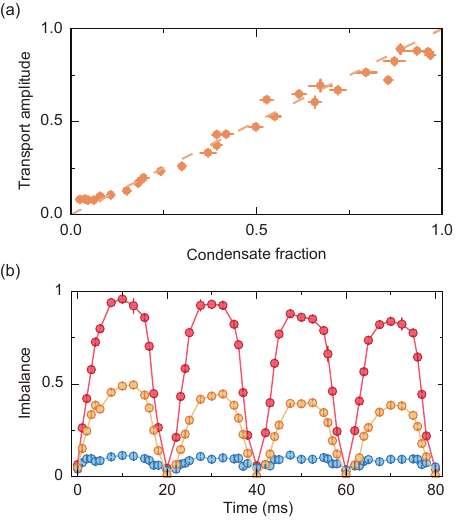}
\caption{Condensate fraction and coherent transport.
(a) Comparison between the condensate fraction measured before lattice loading and the transport amplitude extracted from the population imbalance. The dashed line indicates equal values.
(b) Driven oscillations induced by periodically modulating the energy offset $\Delta(t)$. Red and orange points correspond to condensate fractions of $\eta=0.97(3)$ and $0.50(1)$, respectively, while blue points represent the normal gas. The superfluid component exhibits nearly full-amplitude oscillations, whereas the normal gas shows only a weak response. Error bars denote standard deviations.}
\label{FigS2}
\end{figure}

The relative populations of the condensed and thermal components are controlled by varying the final stage of evaporative cooling.
Before lattice loading, the condensate fraction is extracted from a bimodal fit to the momentum distribution after time-of-flight expansion.
The gas is then adiabatically loaded into the one-dimensional optical lattice, forming an array of coupled quasi-two-dimensional systems.
Although the lattice strongly modifies the momentum distribution along $x$, the transverse momentum distribution remains essentially unaffected~\cite{Chauveau2023,Tao2023}.
We therefore extract the coherent fraction $\eta_{\rm 2D}$ from a bimodal fit to the transverse momentum distribution along $y$.
As shown in Fig.~\ref{FigS1}, the coherent fraction after lattice loading agrees with the condensate fraction of the initial 3D gas over the measured temperature range, with no appreciable heating observed during loading.

We further characterize the coherent component through adiabatic transport.
Starting from balanced odd and even populations, we slowly modulate the energy offset between neighboring wells.
The driving timescale is chosen to be slow compared with coherent tunneling but short compared with normal transport, such that the coherent component follows the modulation while the thermal component remains nearly stationary.
For $V_s=12.2(1)~E_r$, the long lattice is ramped from $V_l=0$ to $2.5(1)~E_r$ within $30~\mathrm{ms}$.
The resulting transport amplitude agrees with the independently measured condensate fraction over a wide temperature range, as shown in Fig.~\ref{FigS2}(a).

As an additional calibration, we periodically drive the energy offset as
$\Delta(t)=\Delta_0[1-\cos(2\pi t/t_0)]/2$, with $t_0=20~\mathrm{ms}$ and $\Delta_0$ larger than the chemical potential.
The coherent component follows the drive with nearly full-amplitude oscillations, whereas the thermal component shows little response on this timescale (Fig.~\ref{FigS2}(b)).
For partially condensed samples, the oscillation amplitude follows the independently measured condensate fraction.
These observations establish the condensate fraction measured from the transverse momentum distribution as an experimental measure of the coherently transporting component in our lattice system.

\section{Measurements of transport rates}

Starting from a balanced configuration, we characterize the transport by varying the total ramp duration $\tau$ at fixed tunneling strength.
For each $\tau$, the energy offset is linearly ramped from its initial to final value, and the populations of the odd and even subsystems are measured at the end of the ramp.
The resulting imbalance therefore probes the accumulated transport during the ramp.
During detection, the lattice depths are rapidly increased to map the site populations onto different Bloch bands.
The finite mapping time introduces additional dynamics and produces a small initial imbalance, which becomes more pronounced at larger tunneling strengths.

For samples dominated by either the superfluid or normal component, the imbalance dynamics are fitted with
$\mathcal{I}=\mathcal{I}_0-A_{s,n}e^{-t/\tau_{s,n}}$,
where $\tau_{s,n}$ denotes the corresponding characteristic transport time.
For samples containing both components, we use
$\mathcal{I}=\mathcal{I}_0-A_s e^{-t/\tau_s}-A_n e^{-t/\tau_n}$,
which resolves the fast superfluid and slow normal transport timescales.
Small oscillations superimposed on the relaxation arise from residual Josephson dynamics.

\begin{figure}[htbp]
\includegraphics[width=7.8cm]{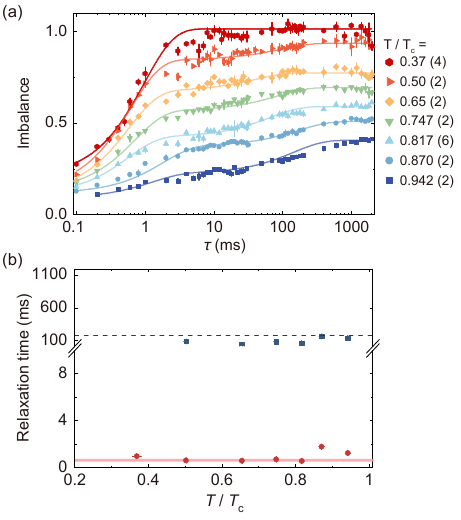}
\caption{Transport dynamics in mixed samples.
(a) Population-imbalance dynamics for $T/T_c$ ranging from $0.37(1)$ to $0.942(1)$, shown on a logarithmic time axis. Solid lines are double-exponential fits, resolving the fast superfluid and slow normal transport.
(b) Extracted relaxation times as a function of $T/T_c$. The two timescales remain well separated and show only weak temperature dependence. The red line denotes the theoretical simulation of the superfluid dynamics, while the blue dashed line indicates the calculated thermalization time for the normal component.}
\label{FigS3}
\end{figure}

The two timescales remain clearly separated throughout the coexistence regime.
As the temperature increases, the amplitude of the fast component decreases while that of the slow component increases, consistent with the changing coherent and thermal fractions (Fig.~\ref{FigS3}(a)).
The extracted $\tau_s$ and $\tau_n$ depend only weakly on temperature at fixed lattice parameters, as shown in Fig.~\ref{FigS3}(b).
This behavior indicates that the characteristic rates are primarily determined by the underlying transport mechanisms rather than by the relative populations of the two components.

\begin{figure*}[hbtp]
\includegraphics[width=13cm]{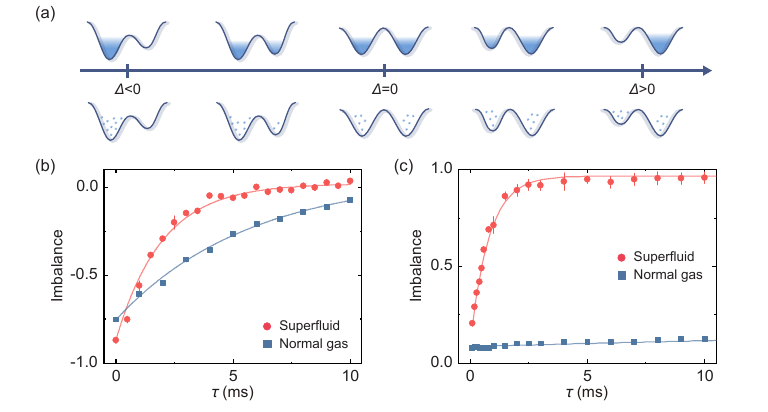}
\caption{Transport dynamics for different initial configurations.
(a) Transport during a gradual ramp of the energy offset $\Delta$. The superfluid continuously follows the changing potential, whereas the normal gas initially fills the empty sites but responds only weakly to further driving after reaching a balanced population.
(b) Transport into initially empty sites. Superfluid and normal samples are initially loaded into the odd sites of isolated double wells. Both components approach a balanced population within $10~\mathrm{ms}$, with relaxation times of $1.97(8)~\mathrm{ms}$ and $5.7(5)~\mathrm{ms}$, respectively.
(c) Transport from initially balanced populations. In the connected double-well array, the superfluid rapidly develops a population imbalance, with a relaxation time of $0.80(3)~\mathrm{ms}$, whereas the normal gas shows little transport on the same timescale.
}
\label{FigS4}
\end{figure*}

The initial population configuration plays an important role in the normal-gas dynamics.
To distinguish single-particle tunneling from the slow transport observed in the main experiment, we prepare isolated double wells with all atoms initially occupying the odd sites.
For the superfluid, the population rapidly transfers to the initially empty even wells through coherent Josephson tunneling.
The characteristic time is slightly longer than in the coupled array because fewer tunneling channels are available.

The normal gas also rapidly expands into initially empty wells, with a timescale comparable to the single-particle tunneling time (Fig.~\ref{FigS4}(b)).
This demonstrates that the slow normal transport observed in the main experiment does not originate from suppressed tunneling itself.
Instead, when the two subsystems are initially populated and thermodynamically balanced, applying an energy offset requires a redistribution of both particles and energy to establish a new population imbalance.
This process is limited by thermalization between the axial lattice motion and the in-plane thermal degrees of freedom, resulting in the much slower dynamics shown in Fig.~\ref{FigS4}(c).
The corresponding thermalization mechanism is discussed in the following section.

\section{Superfluid dynamics}

\begin{figure}[!b]
\includegraphics[width=8.2cm]{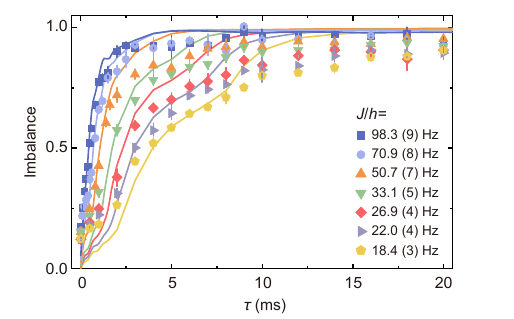}
\caption{Dynamics of the superfluid. The relative temperature of the atomic clouds is $T/T_c = 0.36(1)$. The dots represent experimental data obtained at different tunneling strengths, while the solid lines correspond to simulation results. The theoretical predictions exhibit good agreement with the experimental observations.}
\label{FigS6}
\end{figure}

For a weakly interacting Bose gas, the phase-coherent superfluid component can be described by a macroscopic order parameter $\Psi(\mathbf{r},t)$, whose mean-field dynamics obey the Gross-Pitaevskii (GP) equation~\cite{Pethick2008,Pitaevskii2016}.
In the tight-binding regime, the order parameter can be expanded as
$\Psi(\mathbf{r},t)=\sum_j\psi_j(t)\Phi_j(\mathbf{r})$, where
$\psi_j(t)=\sqrt{N_j(t)}e^{i\phi_j(t)}$ describes the population $N_j$ and phase $\phi_j$ of the superfluid component at site $j$, and $\Phi_j(\mathbf{r})$ is the corresponding localized mode.
Integrating over the transverse degrees of freedom gives the discrete GP equation~\cite{Smerzi2003},
\begin{equation}
\label{eq:Hamiltonian}
i\hbar\frac{\partial\psi_j}{\partial t}
= -J(\psi_{j+1}+\psi_{j-1})
+(\varepsilon_j+\mu_j)\psi_j,
\end{equation}
where $J$ is the nearest-neighbor tunneling energy, $\varepsilon_j$ is the local single-particle energy, and $\mu_j$ accounts for the mean-field interaction.

\begin{figure*}[htbp]
\includegraphics[width=15.5cm]{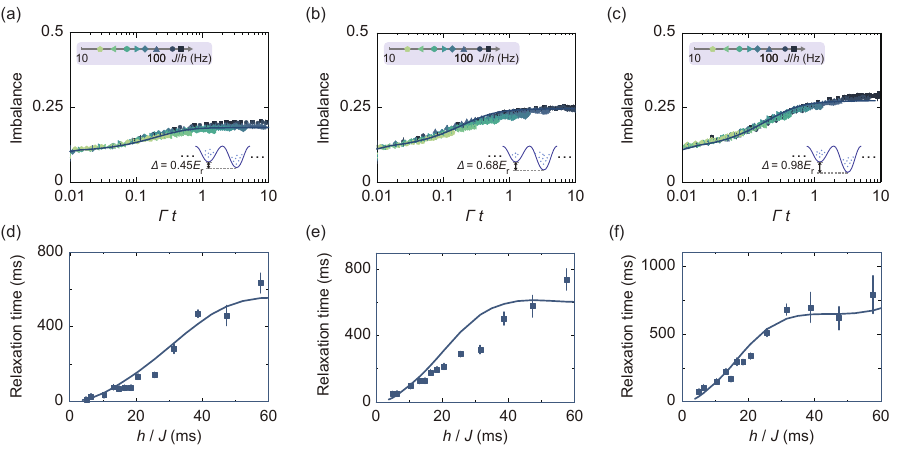}
\caption{
Normal-gas transport at different energy biases.
(a-c) Scaled transport dynamics of the normal gas for $\Delta=0.45E_r$, $0.68E_r$, and $0.98E_r$, respectively.
(d-f) Corresponding characteristic relaxation times for the same energy biases.
Each column shows the dynamics and relaxation times under the same conditions, demonstrating consistent scaling behavior across different energy biases.
Points denote experimental data, and solid lines show the theoretical simulations.
}
\label{FigS7}
\end{figure*}

Writing the GP equation in terms of the site populations and phases gives
\begin{equation}
\label{eq:junctionarray}
\begin{split}
\hbar\dot{N}_{j}
=&\,2J\sqrt{N_jN_{j-1}}\sin(\phi_j-\phi_{j-1})\\
&+2J\sqrt{N_jN_{j+1}}\sin(\phi_j-\phi_{j+1}).
\end{split}
\end{equation}
For nearly balanced neighboring sites, $N_j\simeq N_{j\pm1}$, and approximately symmetric phase differences, this relation reduces to the Josephson current-phase relation.
The maximum flow rate per particle is then $I_c=2J/\hbar$, which sets the characteristic scale of coherent superfluid transport.
Accordingly, the transport dynamics are expected to scale with $Jt/h$, consistent with the collapse observed experimentally in Fig.~\ref{Fig3}(a).

Experimentally, the energy offset $\Delta(t)$ is slowly ramped to drive the superfluid toward the deeper wells.
The resulting chemical-potential difference generates a finite phase difference and hence a Josephson current.
For a sufficiently slow ramp, $\Delta\phi$ remains small and the current stays below the critical value $I_c$.
As shown in Fig.~\ref{Fig3}(a), measurements at different tunneling strengths collapse when time is rescaled by $Jt/h$, confirming that $J$ sets the characteristic scale of the superfluid transport.
For faster driving, coherent oscillations and additional excitations appear, leading to deviations from the scaled dynamics.

We numerically simulate the superfluid dynamics using the GPELab toolbox~\cite{Antoine2015}.
The optical lattice divides the cloud into an array of coupled quasi-2D systems.
Within the local density approximation, the chemical potential at site $j$ is given by
$\mu_j=\mu_c-m\omega_x^2x_j^2/2$, where $x_j$ denotes the site position.
The ground state is obtained self-consistently using the experimentally measured atom number and trapping parameters.
To reduce the computational cost, we assume cylindrical symmetry in the transverse plane and evolve the GP equation under the same long-lattice ramp used in the experiment.
The site populations are extracted from the final density distribution to determine the population imbalance.

We repeat the calculation for different tunneling strengths $J$ by varying the short-lattice depth $V_s$ according to the experimental conditions.
As shown in Fig.~\ref{FigS6}, the simulations quantitatively reproduce the measured superfluid dynamics over the explored range of tunneling strengths.
Small oscillations in the simulated curves arise from residual Josephson dynamics when the ramp is relatively fast.

\begin{figure*}[htbp]
\includegraphics[width=15.5cm]{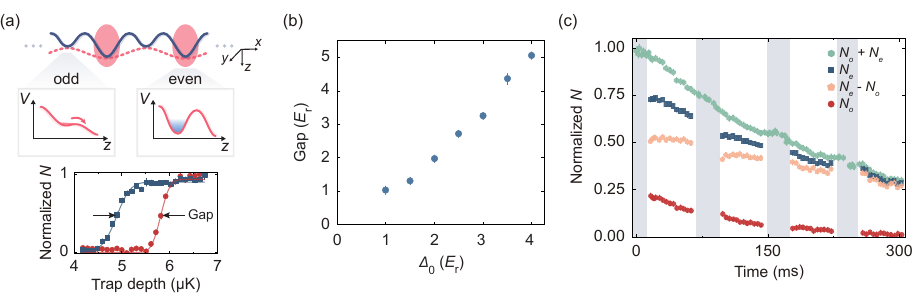}
\caption{Superfluid purification method.
(a) Site-selective removal. The long lattice produces different transverse confinements for the odd and even subsystems. By lowering the additional dipole trap, atoms on the odd sites escape while those on the even sites remain trapped, as shown by the measured atom numbers.
(b) Site-dependent confinement. The difference between the trapping potentials of the two subsystems increases with the long-lattice depth. Two independent measurements show consistent results.
(c) Cyclic purification. Atom-number evolution during repeated separation and removal cycles for an initial condensate fraction of approximately $50\%$. Shaded regions indicate the transport stages. The normal component is progressively filtered out over successive cycles, increasing the condensate fraction to $94(1)\%$ after four cycles.
}
\label{FigS8}
\end{figure*}

\section{Normal-gas transport}

We consider two neighboring quasi-2D normal gases coupled by tunneling.
Initially, the two subsystems have equal populations and a common temperature $T_i$ at zero energy offset.
Applying an offset $\Delta$ raises the odd sites relative to the even sites and drives atoms toward the lower-energy subsystem.
The released potential energy is converted into thermal energy, resulting in a final temperature $T_f>T_i$.
At equilibrium~\cite{Yefsah2011,Zapata1998}, the two subsystems share the same temperature and generalized chemical potential, with
$\mu_{\rm o}+\Delta=\mu_{\rm e}$ and $T_{\rm o}=T_{\rm e}=T_f$.
Thus, the population redistribution is necessarily accompanied by energy redistribution.

The normal transport involves two well-separated dynamical processes.
Particle tunneling occurs at a characteristic rate $J/h\sim10^2~\mathrm{s^{-1}}$.
Elastic collisions dephase the tunneled atoms at a comparable rate,
$n\sigma\bar v_{\rm rel}\sim10^2~\mathrm{s^{-1}}$.
Together, tunneling and dephasing lead to rapid incoherent population exchange between neighboring subsystems, characterized by an effective rate $\gamma_{\rm J}\sim10^2~{\rm s^{-1}}$ in our experiment.
Since the forward and backward processes are nearly balanced, this fast exchange alone produces little net transport.

After tunneling, net redistribution requires the energy released along the
lattice direction to be transferred to the transverse thermal degrees of
freedom. This cross-dimensional thermalization occurs at a rate
$\Gamma_{\rm th}$, which is strongly suppressed in the quasi-2D regime
~\cite{Petrov2001}. Defining $\chi=\hbar\omega_l/(k_{\rm B}T_i)$, the
thermalization rate is
\begin{equation}
\label{eq:thermalization_rate}
\Gamma_{\rm th}
=
\frac{16\Omega}{9\pi}
\frac{\chi e^{-\chi}(1-e^{-\chi})^3}
{1+8\sigma_l^2/(9\pi a_s^2)},
\end{equation}
where $\Omega=\omega_{\perp}^2N_j/\omega_l$. For our experimental parameters,
$\Gamma_{\rm th}$ is typically a few ${\rm s^{-1}}$.

The corresponding particle-redistribution rate is
\begin{equation}
\gamma_{\rm th}=C_{\rm th}\Gamma_{\rm th},
\end{equation}
where
$C_{\rm th}=2(k_{\rm B}T_i/\Delta)
[1/2+4\chi^{-2}\sinh^2(\chi/2)]^{-1}$ accounts for the thermodynamic
conversion between energy relaxation and population redistribution.
For our parameters, $C_{\rm th}$ remains close to unity, with its precise
value depending on the lattice depth.

Following the Landauer description introduced in the main text, the two
subsystems are described by Bose distributions
\begin{equation}
\label{eq:bose_distribution}
f_{\rm o,e}(E,t)
=
\frac{1}
{\exp[(E-\mu_{\rm o,e}(t))/k_{\rm B}T_{\rm o,e}(t)]-1}.
\end{equation}
The energy-dependent transmission is
\begin{equation}
\label{eq:transmission}
\mathcal{T}(E)
=
\frac{\hbar^2\gamma_{\rm J}\gamma_{\rm th}}
{E^2+\hbar^2(\gamma_{\rm J}+\gamma_{\rm th})^2/4}.
\end{equation}
The Bose distributions and transmission determine the instantaneous particle
and energy currents through the Landauer relations given in the main text.

The fast particle exchange and slow thermalization act successively, giving
the normal relaxation rate
\begin{equation}
\label{eq:normal_relaxation}
\Gamma
=
\frac{\gamma_{\rm J}\gamma_{\rm th}}
{\gamma_{\rm J}+\gamma_{\rm th}}
\simeq \gamma_{\rm th},
\qquad
\gamma_{\rm J}\gg\gamma_{\rm th}.
\end{equation}
Thus, in our experimental regime, the normal transport is limited by
cross-dimensional thermalization.

For quantitative comparison, we numerically evolve the two subsystems in short
time steps using the Green-function approach. Spatial inhomogeneity is included
within the local-density approximation. At each step, the populations, chemical
potentials, and temperatures are updated from the calculated particle and energy
currents, and the transmission is recalculated accordingly. This procedure
yields the full imbalance dynamics shown in the main text. Fitting the calculated
curves with the same exponential form used for the experimental data gives
relaxation rates in good agreement with the measurements. As shown in
Fig.~\ref{FigS7}, this agreement and the scaling behavior remain consistent for
different energy biases, further supporting the thermalization-limited nature
of the normal transport.

\section{Superfluid purification}

We purify the superfluid by combining dynamical separation with site-selective atom removal.
The transport duration is chosen to be nearly adiabatic for the superfluid but short compared with the normal transport timescale, such that the superfluid preferentially moves to the even sites while the normal gas remains largely unchanged.
Atoms on the odd sites are then selectively removed.
Repeating this separation and removal progressively reduces the normal component and increases the superfluid fraction.

To selectively remove atoms from the odd sites, we exploit the site-dependent transverse confinement of the superlattice.
The superlattice phase is set to $\varphi=\pi/4$, producing an energy offset $\Delta_0=V_l$ between the odd and even subsystems (Fig.~\ref{FigS8}(a)).
In this configuration, the blue-detuned short lattice provides anti-confinement for the odd sites, whereas the red-detuned long lattice provides additional transverse confinement for the even sites.
By lowering the additional dipole trap, atoms on the odd sites escape under gravity while those on the even sites remain trapped.
This provides a site-selective removal mechanism for the purification protocol.

\begin{figure*}[!htbp]
\includegraphics[width=15.5cm]{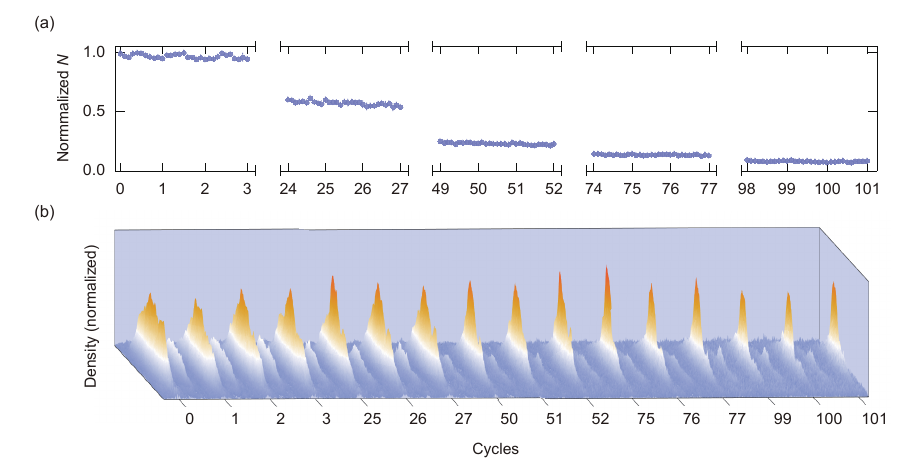}
\caption{Long-lived driven oscillations under purification.
(a) Normalized atom number during the driven dynamics with cyclic purification. The purification continuously removes the thermal component, while coherent oscillations are maintained as the total atom number decreases to approximately $10\%$ of its initial value.
(b) Representative odd-site density profiles at the maximum and minimum populations during the driven oscillation, illustrating the large-amplitude population transfer between the two subsystems.}
\label{FigS10}
\end{figure*}

We characterize the site selectivity by measuring atom loss from the odd and even subsystems as the additional dipole trap is lowered (Fig.~\ref{FigS8}(a)).
For $V_l=4.0(1)~E_r$, atoms on the odd sites are efficiently removed within a $65$-ms hold, while those on the even sites remain largely unaffected.
Fitting the loss curves yields the difference between the escape thresholds of the two subsystems.
Its dependence on the long-lattice depth agrees with the calculated site-dependent confinement, as shown in Fig.~\ref{FigS8}(b).

Separation and removal are combined into a cyclic protocol by modulating the long-lattice depth.
During the first part of each cycle, the increasing energy offset drives the superfluid toward the even sites while the normal component responds only weakly.
The transverse confinement is then lowered to selectively remove atoms remaining on the odd sites.
After removal, the long lattice and dipole trap are returned to their initial values, preparing the system for the next filtering cycle.

Repeating the cycle progressively removes the normal component.
Since the normal gas is approximately equally distributed between the two subsystems and only atoms on the odd sites are removed in each cycle, the remaining normal fraction is expected to decrease approximately as $2^{-n}$ after $n$ cycles.
With four cycles, the measured superfluid fraction increases from approximately $50\%$ to $94(1)\%$ (Fig.~\ref{Fig4}(b) and Fig.~\ref{FigS8}(c)).
The accompanying superfluid loss mainly arises from the limited site selectivity of the removal process at higher temperatures and can be reduced by weaker removal over more cycles and improved trap control.

Rapid changes in the lattice potential can generate excitations, including density fluctuations and phase slips, which dissipate energy and suppress coherent oscillations. We characterize these ramp-induced excitations by measuring the population imbalance after fast lattice ramps, as shown in Fig.~\ref{Fig4}(c). Such excitations progressively reduce the coherent superfluid response during repeated driven dynamics.

We then apply the purification protocol continuously during the driven Josephson dynamics. The tunneling rate is set to $J/h=217(5)~\mathrm{Hz}$, with a driving period of $10~\mathrm{ms}$. Without purification, the oscillation amplitude gradually decays as excitations accumulate. With purification, the generated thermal component is continuously removed, maintaining coherent oscillations for more than $100$ cycles (Fig.~\ref{Fig4}(d)). Remarkably, the oscillations remain clearly visible even after the total atom number decreases to approximately 10\% of its initial value, as further shown in Fig.~\ref{FigS10}. These observations demonstrate that dynamical filtering can continuously remove generated excitations and sustain coherent superfluid dynamics beyond the timescale accessible without purification.

\end{document}